\documentstyle[12pt]{article}
\begin{document}
\thispagestyle{empty}
\large
\vspace*{-5cm}
\Large
\begin{center}
{\bf Determination  of the branching ratios}\\
$\Gamma (K_L \to 3 \pi^0) / \Gamma (K_L \to \pi^+ \pi^- \pi^0)$\\
{\bf and} \\
$\Gamma (K_L \to 3 \pi^0) / \Gamma (K_L \to \pi e \nu )$\\[7ex]
\normalsize
%
%
A.~Kreutz, M.~Holder, M.~Rost, R.~Werthenbach\\
{\it Fachbereich Physik, Universit\"at Gesamthochschule Siegen,
Germany$^{1}$}\\[2ex]
K.J.~Peach\\
{\it Physics Department, University of Edinburgh, United Kingdom}\\[2ex]
H.~Bl\"umer, R.~Heinz, K.~Kleinknecht, P.~Mayer, B.~Panzer$^2$, \\
B.~Renk, H.~Rohrer, A.~Wagner\\
{\it Institut f\"ur Physik, Universit\"at Mainz, Germany$^3$}\\[2ex]
E.~Aug\'e, D.~Fournier, L.~Iconomidou-Fayard, O.~Perdereau, \\
A.C.~Schaffer, L.~Serin\\
{\it Laboratoire de l'Acc\'el\'erateur Lin\'eaire, IN2P3-CNRS, Universit\'e de
Paris-Sud,\\ Orsay, France$^4$} \\[2ex]
L.~Bertanza, A.~Bigi, P.~Calafiura$^5$, M.~Calvetti$^6$, R.~Carosi,
R.~Casali, \\
C.~Cerri, I.~Mannelli$^5$, V.~Marzulli$^5$, A.~Nappi$^{7}$, G.M.~Pierazzini\\
{\it Dipartimento di Fisica e Sezione INFN, Pisa, Italy}\\[5ex]
ABSTRACT\\
\end{center}
\begin{small}
\begin{quote}
Improved branching ratios were measured for the $K_L \to 3 \pi^0 $ decay
in a neutral beam at the CERN SPS with the NA31 detector:\\
$\Gamma (K_L \to 3 \pi^0) / \Gamma (K_L \to \pi^+ \pi^- \pi^0)
= 1.611 \pm 0.037$ and\\
$\Gamma (K_L \to 3 \pi^0) / \Gamma (K_L \to \pi e \nu )
= 0.545 \pm 0.010$.\\
{}From the first number an upper limit
for $\Delta I =5/2$ and $\Delta I = 7/2 $ transitions in neutral
kaon decay is
derived. Using older results for the Ke3/K$\mu $3 fraction, the 3$\pi^0$
branching ratio is found to be
$\Gamma (K_L \to 3 \pi^0 )/ \Gamma_{tot} = (0.211 \pm 0.003)$,
about a factor three more precise than from previous experiments.\\[1ex]
\end{quote}
\end{small}
\scriptsize
$^1$    Funded by the German Federal Minister for Research and
        Technology (BMFT) under contract 054Si74\\
$^2$    Present address: CERN, Geneva, Switzerland\\
$^3$    Funded by the German Federal Minister for Research and
        Technology (BMFT) under contract 054Mz18\\
$^4$    Funded by Institut National de Physique des Particules et de
        Physique Nucl\'eaire (IN2P3), France\\
$^5$    Present address: Scuola Normale Superiore, Pisa, Italy\\
$^6$    Present address: Dipartimento di Fiscia e Sezione INFN, Perugia,
        Italy\\
$^{7}$  Present address: Dipartimento di Scienze Fisiche e Sezione INFN,
        Cagliari, Italy\\[3ex]
\normalsize
\begin{center}
{\sl (To be published in Zeitschrift f\"ur Physik C)}
\end{center}
\clearpage
%
%
%
\noindent
{\large \bf 1. Introduction}\\[1.0ex]

The hadronic decays of particles involving heavy quarks are
not yet well understood. Perhaps the best known example of an
empirical law is the $\Delta I = 1/2$ rule, which states that in strange
particle decays the $\Delta I =1/2$
amplitude dominates over the $\Delta I =3/2 $ amplitude. These are the only
amplitudes available in $K \to 2\pi$ decays. While transitions with $\Delta I
= 5/2$ and $ \Delta I =7/2$ are possible in $K\to 3\pi$ decays, they are not
allowed in the conventional first order weak interaction. The I=3 state of
pions is only accessible through additional
mechanisms like the electromagnetic interaction that do not
conserve isotopic spin.
The best limits for $\Delta I =5/2$ and
$\Delta I =7/2$ transitions in $K \to 3 \pi$ decays were obtained from
charged kaon decays. A similar precision could not be reached in neutral kaon
decays because of the difficulty of measuring the $K_L \to 3 \pi^0$ decay.
Only recently a small, but expected,
deviation from a pure phase-space density
of final states in this decay was observed $^{1)}$. Additional information can
be obtained from a comparison of the branching ratios of $K_L \to 3 \pi^0$
and $K_L \to \pi^+ \pi^- \pi^0$ decays.\par

In this paper new measurements are presented of the branching ratios of
\newline
$K_L \to 3 \pi^0$/$K_L \to \pi^+ \pi^- \pi^0$ and
$K_L \to 3 \pi^0$/$K_L \to \pi e \nu$.
As a byproduct, the absolute branching ratio of the
$K_L \to 3 \pi^0$ decay is determined with better precision.
This decay is convenient for a determination of the kaon flux
in experiments on neutral decays of the $K_L$ at high energies because
it is essentially free of backgrounds from
other decays. The branching ratio of $K_L \to 3 \pi^0/K_L \to all \; charged$
has been measured once $^{2)}$, with a relative uncertainty of 4 $\%$.\par

After a brief description of the apparatus in section 2,
the event selection
and the analysis of these three decays are discussed in section 3.
The results are presented in section 4. \\[4ex]
{\large \bf 2. Apparatus and data taking}\\[1ex]

The detector used for this experiment was designed for a measurement
of direct CP-violation in a high energy neutral kaon beam
(experiment NA31) at the CERN SPS. It is described
in detail in previous publications $^{3,4)}$.
For the benefit of the reader a sketch of the experimental layout is
shown in fig. 1; it does not include the transition radiation detector.
A special run with trigger conditions different from those in the NA31
experiment was made for the purpose of the present experiment.

The detector consists of two multiwire proportional chambers, each with
four planes of different wire orientation. The chambers are separated by
23 m and are followed by a lead/liquid-argon electromagnetic calorimeter,
an iron/scintillator hadron calorimeter and a muon veto counter system.
A transition radiation detector, located between the second chamber and
the calorimeters, is used for additional electron identification.
The lateral division in 1.25 cm wide horizontal and vertical strips
in the electromagnetic calorimeter allows a measurement of shower
position and width. Due to the calorimeter length (25 radiation lengths),
electromagnetic showers rarely leak into the hadron calorimeter.
A scintillator hodoscope in front of the electromagnetic calorimeter
provides trigger and timing information for charged particles;
a scintillator plane in the liquid argon
serves the same purpose for neutral decays. The decay region is surrounded by
four rings of veto counters which are used in the trigger to suppress decays
in which a photon or a charged particle escapes the detector. An
evacuated beam pipe with an average radius of 10 cm
traverses the whole detector
and allows particles in the primary beam (primarily neutrons and
photons) to pass directly to the beam dump and not produce
backgrounds in the detector.
The momentum range of kaons accepted in the analysis is from 70 GeV to 150 GeV.

The trigger for neutral decays requires a coincidence of scintillators in
opposite halves of the electromagnetic calorimeter along with
a total electromagnetic energy of more than 35 GeV, as
well as less than 12 GeV in the hadron calorimeter. With the pulse heights
recorded in the horizontal and vertical strips of the
electromagnetic calorimeter, the number of
peaks in each projection is determined on-line by fast analog
electronics.  At least
five peaks in one projection are required for $3 \pi^0$ decays. The distance
between the decay point and the calorimeter, calculated from the first and
second moments of the photon distributions under the assumption that
all the photons come from the decay of a neutral kaon,
has to exceed 55 m.

Charged decays are selected by a coincidence of hodoscope counters
in opposite quadrants. For these events an energy deposition
of more than 30 GeV
in the calorimeters is required in the trigger. Further cuts then are
imposed on the fraction of events for which on-line reconstruction
is possible. All events with a photon shower more than 15 cm from any
charged particle in either the vertical or horizontal
projection are accepted. In order to suppress Ke3 decays, charged events
without a photon candidate and with two space points in each wire chamber
are "downscaled" and only one of five of these events is recorded.

A total of $1.2 \times 10^6$ triggers were recorded during one day at
the end of the last running-period of NA31. The $K_L$ beam was run at
half the normal intensity.
Before and  after the run the detector was calibrated {\it in situ} with
$K_S \to 2 \pi$ decays from a $K_S$ beam. Details of the data analysis
that follows are given in Ref.5. \\[4ex]
\clearpage
\noindent
{\large \bf 3. Data analysis}\\[2ex]
The data analysis closely follows
the procedure established previous\-ly $^{6,7,8)}$.
For convenience the same fiducial volume for the decay points (between
2 m and 48m from the end of the $K_L$ collimator) was chosen as in the
NA31 experiment. This choice is originally motivated by the extent over which
the $K_S$ target can be moved. For each of the decay modes
the kaon flux is determined - up to a constant factor - as a function of the
kaon energy from the observed data and the calculated acceptance.
The relative branching ratios are then measured in bins of kaon energy.
The final result is averaged over energy. As a convenient way to
describe the kaon energy spectrum, a function of the form
$f(E) \propto p(E) \cdot E^{\alpha} \cdot e^{-\beta E}$
with a second order polynomial $p(E)$
is used in the Monte Carlo programs with parameters adjusted to describe
the $K_L \to 3\pi^0$ data (see fig. 2).
This spectrum describes correctly all three decay rates; it is also
consistent with the spectrum observed in the NA31 experiment from
$K_L \to 2\pi^0$ decays in the same beam.
A precise description of the spectrum is not
crucial; what matters in this experiment is the ratio of
acceptances.
The pertinent points of the analysis are given in the following
sections.\\[2ex]
{\large \bf 3.1 $K_L \to 3 \pi^0$ decays}\\

Events with at least 6 reconstructed photons whose energies are greater
than 5 GeV and which are separated
from each other by a minimum distance of 5 cm (3 cm in both x- and
y-projections if they fall in the same quadrant) are retained for further
analysis. Each photon is required to be at least 16 cm from the center of
the beam line to avoid energy leakage into the beam pipe. The longitudinal
position of the decay point is reconstructed using the kaon invariant mass
and is required to be between 2 m and 48 m downstream of the final
$K_L$ collimator.
The best pairing of the photons to $3\pi^0$s is chosen to determine the
vertex and momentum of the kaon, but no cuts on
the 2-photon invariant masses are made. No cut is made on the center of
gravity of the photon showers, since only
0.08\% of events have a center of gravity in excess of 10 cm. This is
well outside the edge of the beam profile at 5 cm from the beam
centerline. The energy scale is calibrated with 0.1$\%$ relative
precision in $K_S$ runs using the known position of an anticounter at the
exit of the $K_S$ collimator. The systematic difference between calibrations
before and after the $K_L$ run is below $0.2\%$.
The energy scale is consistent within $(0.2 \pm 0.1) \% $
with the scale deduced from $K_L\to\pi^+\pi^-\pi^0$ decays, in which the
calculated decay point of the $\pi^0$ can be compared with the measured
intersection of $ \pi^+$ and $ \pi^-$ tracks. The total kaon energy is
confined to the interval from 70 GeV to 150 GeV.

The acceptance is calculated for a uniform density in the
Dalitz plot. The quadratic term does not contribute a statistically
significant effect to the acceptance.  The simulations of
the photon energy and impact point fluctuations are made
with parameters determined in the NA31 experiment $^{3)}$. The measured
distributions of $\pi^0$ masses, photon energies and impact points
agree well with the simulation (see, e.g., the photon energy distribution
shown in fig. 3).  The acceptance depends on the kaon energy (fig. 4).
It is $(6.55\pm 0.07) \% $ on average for the decay $K_L \to 3 \pi^0$.
The uncertainty reflects the variation
in the data with changing cuts. Most sensitive are the cuts in
the minimum photon energy (varied from 3 GeV to 7 GeV) and the projected
distance between photons (varied from 2 cm to 4 cm).
The relative change of the acceptance is 0.7\% and 0.8\%, respectively.
The spectrum obtained from
$K_L \to 3 \pi^0$ decays is compared in fig. 5 with the Monte Carlo
simulation. The agreement is satisfactory. This is not surprising, since
the parameters of the Monte Carlo are adjusted to describe this spectrum.
The position of the decay point along the the beam (fig. 6) has an almost
uniform distribution, well simulated by Monte Carlo.

Corrections are necessary for photon conversion, $\pi^0$ Dalitz decays,
for accidentals and for reconstruction losses. They are summarized in
table 1.  Events with space points
reconstructed in the first (upstream) wire chamber are not accepted.
The material in front of the first chamber is equivalent to 0.0048
radiation lengths, essentially given by the thickness of
the large vacuum window on the downstream end of the evacuated decay
region.  Accidental tracks in the first wire chamber are distinguished
from conversions on the basis of their projected distance from showers
in the calorimeter. Accidental photons are determined from the
observed rate of 7$\gamma$ events.
In 0.38\% of 6$\gamma$ events an additional photon with more than 2 GeV is
found; in 50\% of these cases the photon has an energy exceeding
5 GeV. From overlays of observed 5$\gamma$ events and 6$\gamma$ events
with these accidental photons we estimate that 0.11\% of 6$\gamma$ events
are lost and 0.08\% are gained by accidental activity.
Reconstruction losses are estimated from a sample of 6768 Monte Carlo
events which pass all cuts and in which full shower profiles as measured
in the detector are simulated. \\[4ex]
{\large \bf 3.2 $K_L \to  \pi e \nu $ decays}\\

Events with exactly two reconstructed space points in the first chamber
and at least two space points in the second chamber are candidates
for Ke3 decays. Cuts in the distance of a track from the center of the
beam pipe in the second chamber (18 cm) and in the position of the
decay point (2m $\le$ z $\le$ 48 m) are imposed.
Electrons are identified by their pulse height
in the transition radiation detector (TRD) and by the small amount of
energy deposited in the hadron calorimeter. The mean of the three lowest
pulseheights from the four proportional chambers of the TRD has to exceed
720 counts.
(For comparison: the mean pulseheights are 460 counts for pions and
1270 counts for electrons.)
This limit is slightly energy dependent to assure a constant
electron identification ($(95.3\pm0.2)\%$) for all particle momenta.
In addition, the energy
deposited in the hadron calorimeter has to be less than 10 $\%$ of the total
particle energy. This cut is $(98.8\pm0.2)\%$ efficient for electrons.
A combined cut is therefore $(94.2\pm0.3)\%$ efficient for electrons; it
reduces
the chance for a pion to be accepted as an electron to about
$0.3\%$. Since the branching ratios of decays without an electron
are smaller than the Ke3 branching ratio, the contamination from other
decays in the Ke3 sample is less than $0.1\%$.
The cuts do not make use of the normally narrow shower width of the
electron; radiative $Ke3$ events are therefore accepted.
Only events with more than one visible photon are rejected.
The losses due to accidental photons are negligible.

Events are further restricted by kinematical cuts. The measured energies
of pions and electrons have to exceed 15 GeV, the pion to electron
energy ratio is required to be
between 0.4 and 2.5, and their sum must exceed 40 GeV.
The calibration of pion energies followes the established NA31 procedure,
which relies on calibrations with pion beams and regular updates with muon
data.
This calibration is further refined in the present experiment with
data from the $K_S$ runs by the
requirement that the invariant mass of the charged pions is centered on the
$K^0$ mass.
The reconstruction of the kaon energy in Ke3 decays
is not possible event by event
because of the well-known ambiguity in the longitudinal neutrino
momentum. The kaon spectrum is determined by an
iterative procedure with appropriate weights for both solutions.
The procedure was tested with simulated events; it reproduces the known
kaon energy spectrum correctly over the kaon energy interval from
70 GeV to 150 GeV. The algorithm is described in the Appendix. The
agreement between the kaon spectrum from Ke3 decays with the Monte Carlo
is reasonable (see fig. 7).
The direct comparison of the $K_L \to \pi e \nu$ and $K_L \to 3\pi^0$
spectra shows even better agreement; the ratio of the decay rates is
consistent with being energy independent (see fig. 12).
The decay point distribution for data and Monte Carlo is shown in fig. 8.

Corrections are necessary for wire chamber inefficiencies, for accidental
tracks and $\delta$-rays in the first wire chamber, for pion punch-through
and for trigger losses (see table 1).
Inefficiencies in the wire chambers, caused by missing signals or by
failure of the reconstruction program in case of additional activity,
are determined from Ke3 events defined by the patterns
of energy deposition in the calorimeters and a clean
signature (only 2 spacepoints) in one of the drift chambers.
The event losses are
$(0.06\pm0.03)\%$ in the first chamber and $(0.16\pm0.04)\%$ in the
second chamber yielding a combined chamber loss of $(0.22\pm0.05)\%$.
Accidentals and $\delta$-rays can give rise to extra hits in the
chambers. In the first chamber, however, events with more than two
hits are rejected.
Losses by $\delta$-rays and accidentals are determined from a sample of
Ke3 events which show two tracks with a common vertex in the fiducial
volume and in which no restriction on the number of space points
in the first chamber is applied.
Pion punch-through, that is muons originating from a pion shower in the
calorimeter and hitting the muon veto counter, is determined from events
taken without a muon veto. Trigger losses, essentially in the on-line
processors, are determined with a sample of events in which
the on-line trigger is avoided but the trigger decision is recorded.

The acceptance is determined by a Monte Carlo simulation of the
$K_L$ decay and of the detector performance. In the simulation a standard
Dalitz-plot density$^{9)}$ is assumed; pion decay-in-flight is included.
The energy resolution of the detector is parametrized by
$\sigma = 0.65*\sqrt{E}$
for pions and $ \sigma = 0.075*\sqrt{E}$ for electrons.
A change of the hadronic
resolution function to $0.72*\sqrt{E}$ causes a relative change of the
acceptance by $0.2 \%$. The estimated uncertainty in the pion energy scale
of 0.3 $\%$ also gives a $0.2\%$ uncertainty in the acceptance.
The main error in the acceptance comes from the cut in the sum of electron and
pion energies (varied from 40 GeV to 50 GeV) and in the ratio of these energies
(varied from 2.5 to 2.2). Both variations causes a relative change of 0.6\%
in the acceptance. A variation in the vertex cut from 2 m to 10 m changes the
acceptance by 0.5\% of its value.
The average acceptance is $(31.93\pm0.35)\%$.\\ [4ex]
{\large \bf 3.3 $K_L \to  \pi^+ \pi^- \pi^0 $ decays}\\

Events are selected which have
two tracks satisfying the same geometric conditions as
described above for Ke3 decays and with two photons of more than 5 GeV
in the calorimeter.
The tracks are required to have no electron
signature in the combination of TRD and calorimeters with the
same definition of electrons as described in section 3.2.
About 6 $\%$ of electrons are not recognized in this way. Contamination
from Ke3 decays, however, is reduced to below
the $0.1 \%$ level because there are two photons in the $\pi^+ \pi^- \pi^0$
final state. The photons must be separated
from the charged particles by more than 20 cm (5 cm in x- and
y-projections) in order to reduce the overlap of
hadron and photon showers in the electromagnetic calorimeter.
One additional photon of less than 5 GeV is allowed.
Losses and gains of events due to accidental photons are estimated by
overlays with accidental photons found in the $3 \pi^0$ sample.
The energies of the charged pions have to exceed 15 GeV and
are further restricted by cuts in
the energy ratio $ (0.4 \le E_1/E_2 \le 2.5)$
and the energy sum $( E_1 + E_2 > 40 $ GeV).
The energies are originally measured by calorimetry. The accurracy of this
measurement is improved by imposing transverse momentum balance in the decay.
The location of the center of gravity
of the pions on the line joining their impact points is better
determined when the impact points of the $\pi^0$ and kaon are used.
The improvement in accuracy of the pion energies is a factor 2 on
average. The kaon momentum spectrum
and the decay point distribution are shown in fig. 9 and 10.
The selected sample is free of background (see fig. 11);
the invariant mass distribution is slightly asymmetric because of
residual overlap of tails from photon and pion showers.

Corrections for accidentals, $\delta$-rays, Dalitz decays, pion
punch-through, trigger losses and photon conversions (see table 1)
are made as discussed in sections 3.1 and 3.2. The
acceptance is calculated with a Monte Carlo program using the known
Dalitz-plot density $^{9)}$.
Systematic uncertainties in the acceptance
are estimated as before from the variations with changes in the cuts.
Most sensitive are the cuts in the minimal radial distance of the
pion trajectories to the beam line in the second drift chamber
(varied from 18 cm to 22 cm) and in the minimal photon energy
(varied from 3 GeV to 7 GeV). They cause a relative change of the acceptance
of 0.8\% and 1.0\%, respectively.
The average acceptance is $(8.64\pm0.16)\%$.\\[4ex]
{\large \bf 4. Results}\\

Among $1.2 \times 10^6$ triggers, 38403 events with $3\pi^0$, 28035
$\pi^+ \pi^- \pi^0$ events and 85520 $\pi e \nu $ events satisfy the
selection criteria. To calculate the number of kaon decays,
the downscaling procedure has to be taken into account.
Since only a fraction of the charged decays is downscaled,
an effective downscaling factor has to be
determined. It is 3.899 $\pm$ 0.008 for Ke3 events and 1.104 $\pm$ 0.0005 for
$\pi^+\pi^-\pi^0$ events. Branching ratios are determined in bins of 10 GeV
for the kaon energy range 70 GeV $< E_K <$ 150 GeV. No systematic
variations with energy are found (see fig. 12, 13 and 14
and table 2). The branching ratios are therefore averaged with the
results:
\begin{eqnarray}
& & \nonumber \\
\frac{\Gamma ( K_L \to 3 \pi^0)}{\Gamma(K_L \to \pi e \nu)}  &=&
      0.545\pm0.004(stat.)\pm0.009(syst.)\\
&=&0.545 \pm 0.010\nonumber \\
& & \nonumber \\
\frac{\Gamma ( K_L \to 3 \pi^0)}{\Gamma(K_L \to \pi^+ \pi^- \pi^0)} &=&
      1.611\pm0.014(stat.)\pm0.034(syst.)\\
&=&1.611 \pm 0.037\nonumber \\
& & \nonumber \\
\frac{\Gamma ( K_L \to \pi^+ \pi^- \pi^0)}{\Gamma(K_L \to \pi e \nu)} &=&
      0.336\pm0.003(stat.)\pm0.007(syst.)\\
&=&0.336 \pm 0.008 \nonumber
\end{eqnarray}
The dominant errors are systematic uncertainties mainly in the
acceptance determination; correlations due to common cuts are taken into
account.
For the final result statistical and systematic
errors are added in quadrature.
\clearpage
\noindent
Using the known
branching ratios $^{9)}$\\[1ex]
$\Gamma(K_L \to \pi \mu \nu)/(\Gamma(K_L \to \pi e \nu) =
0.697 \pm 0.010$ ,\\
$\Gamma(K_L \to 2 \pi)/\Gamma_{tot}=(2.94 \pm 0.05) \times 10^{-3},\\
\Gamma(K_L \to 2\gamma)/\Gamma_{tot}=(0.57 \pm 0.027) \times 10^{-3}$
we obtain
\begin{eqnarray}
  \frac{\Gamma ( K_L \to 3 \pi^0)}{\Gamma_{tot}} &=&
\frac{\Gamma ( 3 \pi^0)}{\Gamma(3 \pi^0)+ \Gamma( \pi^+ \pi^- \pi^0)
      +\Gamma(Ke3)(1+ \frac{\Gamma(K \mu 3)}{\Gamma(Ke3)}) +\Gamma(others)} \\
  & = &      0.2105 \pm 0.0028,
\end{eqnarray}

\noindent
where $\Gamma(others)=\Gamma(K_L \to 2\pi)+\Gamma(K_L \to 2\gamma)$.

With only constant terms in the Dalitz-plot density and no contribution
from $\Delta I=5/2$ and $\Delta I=7/2$ transitions the ratio (2) is expected
to be 1.5. Phase space, Coulomb corrections $^{10)}$, linear $^{11)}$ and
quadratic $^{1,11)}$ terms in the Dalitz-plot density change this number to
$(1.586\pm0.004)$.
If the measured ratio deviates from this value, $\Delta I =5/2$ and
$\Delta I = 7/2$ amplitudes to the $3 \pi$ final state with I=3 are not
negligible. We obtain, in the notation of Ref.10,
\begin{equation}
 \frac { Re \left\{ e^{i \delta_{31}} ( m_{35} - \frac{4}{3} m_{37})
\right\} }
    {  m_{11}- 2 m_{13}}  < 0.024
\end{equation}
with 90 $\%$ confidence level.
Here $m_{11}$ and $m_{13}$ are the amplitudes of the $\Delta I =1/2$
and
$\Delta I = 3/2$ transitions to the I=1 state , and $\delta_{31}$ is the
difference of the s-wave scattering phase shifts between the I=3 and
the symmetric I=1 states of three pions.
A similar limit, on a different combination of $\Delta I = 5/2$
and $\Delta I=7/2$ amplitudes, was obtained $^{10)}$
previously from $3\pi$-decays of charged kaons.\\[6ex]
\clearpage
\thispagestyle{empty}
\begin{table}[h]
\label{SYSALL}
\begin{center}
\caption{\it Gains and losses and systematic uncertainties}
\begin{tabular}{|l|c|c|}\hline
                    & losses in \%         & uncertainty in \% \\ \hline \hline
\multicolumn{3}{|c|}{}                           \\
\multicolumn{3}{|c|}{$K_L\to 3\pi^0$}             \\
\multicolumn{3}{|c|}{}                           \\ \hline
Dalitz-decays $\pi^0 \to e^+e^-\gamma$     & 3.68 & 0.10 \\ \hline
$\gamma$ conversion                        & 2.43 & 0.60 \\ \hline
accidental tracks in first drift chamber   & 0.68 & 0.20 \\ \hline
accidental photons in e.m. calorimeter     & 0.03 & 0.10 \\ \hline
trigger inefficiency                       & 0.03 & 0.03 \\ \hline
reconstruction inefficiency                & 0.19 & 0.05 \\ \hline
                                           &\multicolumn{2}{c|}{}\\
correction factor for $K_L\to 3\pi^0$ &\multicolumn{2}{c|}{1.072$\pm$0.007}\\
                                     &\multicolumn{2}{c|}{}\\ \hline\hline
\multicolumn{3}{|c|}{}                           \\
\multicolumn{3}{|c|}{$K_L\to \pi e \nu$}         \\
\multicolumn{3}{|c|}{}                           \\ \hline
$\delta$-electrons and accidental          &      &      \\
tracks in first drift chamber              & 4.3  & 0.2  \\ \hline
electron-pion identification               & 6.2  & 0.3  \\ \hline
pion punch-through                         & 0.4  & 0.2  \\ \hline
trigger inefficiency                       & 0.39 & 0.24 \\ \hline
wire chamber inefficiency                  & 0.22 & 0.05 \\ \hline
                                     &\multicolumn{2}{c|}{}\\
correction factor for $K_L\to \pi e \nu$&\multicolumn{2}{c|}{1.119$\pm$0.005}\\
                                     &\multicolumn{2}{c|}{}\\ \hline\hline
\multicolumn{3}{|c|}{}                          \\
\multicolumn{3}{|c|}{$K_L\to \pi^+\pi^-\pi^0$}  \\
\multicolumn{3}{|c|}{}                          \\ \hline
$\delta$-electrons and accidental          &      &      \\
tracks in first drift chamber              & 4.3  & 0.2  \\ \hline
electron-pion identification               & 0.2  & 0.1  \\ \hline
Dalitz-decays $\pi^0 \to e^+e^-\gamma$     & 1.20 & 0.03 \\ \hline
$\gamma$ conversion                        & 0.81 & 0.2  \\ \hline
accidental photons in e.m. calorimeter     & 0.09 & 0.10 \\ \hline
pion punch-through                         & 0.8  & 0.4  \\ \hline
trigger inefficiency                       & 0.38 & 0.24 \\ \hline
wire chamber inefficiency                  & 0.22 & 0.05 \\ \hline
                                     &\multicolumn{2}{c|}{}\\
correction factor for $K_L\to \pi^+\pi^-\pi^0$
                                     &\multicolumn{2}{c|}{1.082$\pm$0.006}\\
                                     &\multicolumn{2}{c|}{}\\ \hline
\end{tabular}
\end{center}
\end{table}

\thispagestyle{empty}
\begin{table}[h]
\label{brarat}
\begin{center}
\caption{\it Uncorrected branching ratios with statistical errors}
\begin{tabular}{|c|c|c|c|} \hline
        &                               &             & \\
Energy  & $\frac{\Gamma(K_L \to 3\pi^0)}
                {\Gamma(K_L \to \pi  e \nu)}     $
        & $\frac{\Gamma(K_L \to 3\pi^0)}
                {\Gamma(K_L \to \pi^+\pi^-\pi^0)}$
        & $\frac{\Gamma(K_L \to \pi^+\pi^-\pi^0)}
                {\Gamma(K_L \to \pi  e \nu)}     $ \\
in GeV  &                               &             & \\ \hline \hline
 70 -  80 & 0.558 $\pm$ 0.009 & 1.613 $\pm$ 0.030 & 0.346 $\pm$ 0.005\\
 80 -  90 & 0.579 $\pm$ 0.009 & 1.659 $\pm$ 0.030 & 0.349 $\pm$ 0.005\\
 90 - 100 & 0.564 $\pm$ 0.009 & 1.686 $\pm$ 0.033 & 0.334 $\pm$ 0.006\\
100 - 110 & 0.579 $\pm$ 0.010 & 1.642 $\pm$ 0.036 & 0.353 $\pm$ 0.007\\
110 - 120 & 0.561 $\pm$ 0.011 & 1.491 $\pm$ 0.040 & 0.377 $\pm$ 0.010\\
120 - 130 & 0.567 $\pm$ 0.013 & 1.607 $\pm$ 0.055 & 0.353 $\pm$ 0.012\\
130 - 140 & 0.575 $\pm$ 0.015 & 1.703 $\pm$ 0.078 & 0.338 $\pm$ 0.015\\
140 - 150 & 0.574 $\pm$ 0.018 & 1.539 $\pm$ 0.094 & 0.373 $\pm$ 0.022\\ \hline
\hline
        &          &          &          \\
 70 - 150 & 0.569 $\pm$ 0.004 & 1.626 $\pm$ 0.014 & 0.348 $\pm$ 0.003 \\
        &          &          &          \\ \hline
\end{tabular}
\end{center}
\end{table}

\clearpage
{\large \bf Acknowledgements}\\[3ex]

We thank the members of the CERN NA31 group for their help in running
the beam and the apparatus for this experiment and for numerous suggestions
and discussions during the analysis. We express our gratitude to all the
technical collaborators from the participating institutes
for their continuous effort in the operation of the experiment.
\clearpage

\noindent
{\large \bf Appendix}\\[1ex]

The determination of the $K^0$ spectrum in $K_L \to \pi e \nu $
decays is complicated by the absence of kinematic information on the
neutrino. Therefore, an iterative procedure
to find the true spectrum is applied$^{12)}$.
If $\vec{p}_{e \pi}$ and $E_{e \pi}$ are the measured momentum and energy
of the $( e \pi) $-system in the laboratory,
\begin{eqnarray*}
                     \vec{p}_{e \pi} &=& \vec{p}_e + \vec{p}_{\pi} \\
                     E_{e \pi}       &=& E_e + E_{\pi}
\end{eqnarray*}
with invariant mass
\[
m^2_{e \pi} = E_{e \pi}^2 - \vec{p}^{\;2}_{e \pi},
\]
the center of mass momentum of this system is
\begin{eqnarray*}
      p_{e \pi}^{*^2} &=& E_{e \pi}^{*^2}- m_{e \pi}^2 \\
  &=& \left( \frac{m_K^2 +m_{e \pi}^2}{2m_K}\right)^2 - m_{e \pi}^2 \\
                   &=& \left( \frac{m_K^2 - m_{e \pi}^2}{2m_K}\right)^2 .
\end{eqnarray*}
Since $p_{e \pi}^{*^2} = p_{\nu}^{*^2} $, the total neutrino momentum is known,
and by momentum conservation also the transverse momentum $p_{\perp}$ relative
to the $K_L$ direction, but
for the longitudinal momentum there is an ambiguity:
\begin{displaymath}
   p_l^{\nu}   = \pm \sqrt{p_{\nu}^{*^2} - p_{\perp}^2}.
\end{displaymath}
The two solutions give two possibilities for the parent kaon energy,
$E^+$ and $E^-$.
Let E be the true kaon energy, N(E) the probability density of the
spectrum, and $n(E^+,E^-)$ the observed density of events.
All decays with true energy in the interval (E,E+dE) will end up in two bands
in the $(E^+,E^-)$ space (see fig. 15), with the band limited by the lines
\begin{displaymath}
  (E,E^-),(E+dE,E^-), \qquad E^- < E ,
\end{displaymath}
if the neutrino goes forward, and in the band
\begin{displaymath}
  (E^+,E),(E^+,E+dE), \qquad E < E^+  ,
\end{displaymath}
if the neutrino goes backward in the center of mass system.
Let $w^+$ be the probability for the former case , and  $w^-$ the probability
for the latter case, with
\begin{displaymath}
 \int_0^E w^+(E,E^-) dE^- + \int_E^\infty w^-(E^+,E) dE^+  =1.
\end{displaymath}
The observed density $ n(E^+,E^-)$ is obviously due to decays with $E=E^+$
and $E=E^-$,
\begin{displaymath}
  n(E^+,E^-) = N(E^+) w^+(E^+,E^-) + N(E^-) w^-(E^+,E^-)  ,
\end{displaymath}
 or, divided by $ n(E^+,E^-)$:
\begin{displaymath}
   1= g^+(E^+,E^-) + g^-(E^+,E^-),
\end{displaymath}
where $g^+(g^-)$ denote the probabilities that $E^+(E^-) $ is the correct
solution for the kaon energy. The kaon spectrum is obtained from the
integral equation
\begin{equation}
  N(E)= \int_0^E n(E,E^-)g^+(E,E^-) dE^- +
        \int_E^\infty n(E^+,E)g^-(E^+,E) dE^+ .
\end{equation}
The probabilities $w^+(E,E^-)$ and $w^-(E^+,E)$ are determined from a
simulation of the decays and the detector response. Because of measurement
errors it is  possible to obtain transverse neutrino momenta beyond
the kinematical limit. In these  cases the event is treated as being right at
the kinematical limit, with the consequence $ E^+=E^-$. The solution of
(7) is then obtained by insertion of  the \newline
n-th iteration of the spectrum
on the right hand side to obtain the (n+1)st iteration. Convergence is
reached after about 10 steps. Data are finally corrected for
a $2\%$ scale change in the
simulation between input and output spectra
due to the finite energy resolution of the detector.
\clearpage
{\large \bf References}\\[1ex]
\begin{enumerate}
\item S.V. Somalvar et al., {\it Phys. Rev. Lett.} {\bf 68}(1992) 2580
\item I.A. Budagov et al., {\it Nuovo Cimento} {\bf A57}(1968) 182
\item H.Burkhardt et al., {\it Nucl. Instr. Meth.} {\bf A268}(1988) 116
\item G.D.Barr et al., {\it Nucl. Instr. Meth.} {\bf A294}(1990) 465
\item A. Kreutz, Ph.D.Thesis, University of Siegen, 1993
\item H.Burkhardt et al., {\it Phys. Lett.} {\bf B206}(1988) 169
\item H.Burkhardt et al., {\it Phys. Lett.} {\bf B199}(1987) 139
\item G.D.Barr et al., {\it Phys. Lett.} {\bf B317}(1993) 233
\item Particle Data Group, {\it Phys. Rev.} {\bf D45}(1992), S1
\item T.J.Devlin and J.O.Dickey, {\it Rev. Mod. Phys.}{\bf51}(1979) 237
\item R. Messner et al., {\it Phys. Rev. Lett.} {\bf 33}(1974) 1458
\item A. B\"ohm et al., {\it Nucl. Phys.} {\bf B9}(1969) 605
\end{enumerate}
\clearpage
{\large \bf Figure captions}\\[1ex]
\begin{enumerate}
\item The experimental layout; the transition radiation detector
      is not included.
\item Spectrum of kaons at the target as determined from $K_L \to 3 \pi^0$
      decays. Events with $70 GeV \leq E_K \leq 150 GeV$ are accepted.
\item Distribution of minimum photon energy, $E_{\gamma}$, in
      $K_L \to 3 \pi^0$ events.\\
      Events with $E_{\gamma} > 5 GeV $ are accepted.
\item Detector acceptance for three decay channels.
\item Measured kaon energy distribution of $K_L \to 3\pi^0$ events, not
      corrected for acceptance and decay probability.
\item Vertex distribution of $K_L \to 3\pi^0$ events.
\item Reconstructed kaon energy distribution of $K_L \to \pi e \nu$ events, not
      corrected for acceptance and decay probability.
\item Vertex distribution of $K_L \to \pi e \nu$ events.
\item Measured kaon energy distribution of $K_L \to \pi^+\pi^-\pi^0$ events,
not
      corrected for acceptance and decay probability.
\item Vertex distribution of $K_L \to \pi^+\pi^-\pi^0$ events.
\item Invariant mass distribution of
      accepted $K_L \to \pi^+ \pi^- \pi^0$ events.
\item Ratio of decay rates
      $ \Gamma (K_L \to 3 \pi^0) /\Gamma( K_L \to \pi e \nu )$
      as a function of kaon energy. The shaded area gives the final
      result including the systematic uncertainty.
\item Ratio of decay rates
      $ \Gamma (K_L \to 3 \pi^0) /\Gamma( K_L \to \pi^+\pi^-\pi^0)$
      as a function of kaon energy. The shaded area gives the final
      result including the systematic uncertainty.
\item Ratio of decay rates
      $ \Gamma (K_L \to \pi^+\pi^-\pi^0) /\Gamma( K_L \to \pi e \nu )$
      as a function of kaon energy. The shaded area gives the final
      result including the systematic uncertainty.
\item The two bands in which an event with true energy in (E, E+dE)\break
      will appear (see text).
\end{enumerate}
\clearpage
{\large \bf Table captions}\\[1ex]
\begin{enumerate}
\item (no caption)
\item To obtain the final result, these branching ratios have to be
multiplied by the correction factors in table 1.
\end{enumerate}
\end{document}